\documentclass[11pt, a4paper]{article}
\usepackage{graphicx, color} 
\usepackage[hidelinks]{hyperref}
\usepackage[backend=biber,sorting=none,bibstyle=ieee,citestyle=numeric-comp]{biblatex}
\usepackage{amsmath}
\numberwithin{equation}{section}
\usepackage{slashed}
\usepackage{xurl}
\usepackage{breakurl}
\usepackage{mathtools}
\usepackage{physics}
\usepackage{amssymb}
\usepackage{graphicx}
\usepackage{caption}
\usepackage{mathtools}
\graphicspath{ {./images/} }
\usepackage{csquotes}
\usepackage{newtxtext}
\usepackage{tikz}
\usepackage{enumitem}
\usepackage{tabularx}
\usepackage{array}
\usepackage{lipsum}
\usetikzlibrary{arrows.meta, chains, positioning}
\usepackage{comment}
\usepackage{pgfplots}
\usetikzlibrary{patterns}
\pgfplotsset{compat=1.18} 
\usepackage{pgfplots}
\usepgfplotslibrary{fillbetween}
\usetikzlibrary{decorations.pathreplacing,calligraphy}

\usepackage{authblk}

\definecolor{babyblue}{rgb}{0.63, 0.79, 0.95}

\title{\bf Quantum Gravitational Hair in Gravastars and Observational Tests}

\author[1]{Italo Perrucci \thanks{Email: italo.perrucci@studio.unibo.it}}
\author[2]{Folkert Kuipers\thanks{Email: f.kuipers@physik.uni-muenchen.de}}
\author[1,3]{Roberto Casadio\thanks{Email: casadio@bo.infn.it }}

\affil[1]{\em Dipartimento di Fisica e Astronomia
\protect
\\
	Universit\`a di Bologna
	 \protect\\
	 via Irnerio 46, I-40126 Bologna, Italy}

\affil[2]{\em Arnold Sommerfeld Center for Theoretical Physics \protect\\
	Ludwig-Maximilians-Universit\"at M\"unchen \protect\\
	Theresienstr. 37, 80333 M\"unchen, Germany}

\affil[3]{\em I.N.F.N., Sezione di Bologna \protect\\
	IS - FLAG \protect\\
	via B.~Pichat~6/2, I-40127 Bologna, Italy}

\date{\today}

\begin{document}

\maketitle

\begin{abstract}
Using the effective field theory of quantum gravity at second order in curvature, we calculate quantum corrections to the metric of gravastars
and the closely related dark energy stars. We find that the quantum corrections in the exterior region depend on the equation of state of the
gravastar, thus providing an example of quantum gravitational hair.
We continue by calculating the induced quantum corrections to the photon sphere and the bending of light rays in the weak field regime.
These corrections, albeit Planck scale suppressed, allow in principle to distinguish these objects from black holes observationally. 
\end{abstract}

\section{Introduction}
The final state reached by stars with a mass larger than the Tolman-Oppenheimer-Volkoff limit at the end of the gravitational collapse is generally expected to be a black hole. However, black holes have several peculiar properties that are not yet fully understood. For example, the presence of the singularity indicates a breakdown of general relativity, and it is generally assumed that this singularity will be resolved by a theory of quantum gravity. Furthermore, the presence of an event horizon causally separates the exterior from the interior at the classical level. However, when quantum effects are taken into account, this feature becomes problematic giving rise to the 
information paradox \cite{information_paradox1,information_paradox2}.
Due to these and other paradoxical properties of black holes, significant effort has been devoted to exploring potential alternative models that could describe the final state of a gravitationally collapsed object, while remaining consistent with current observational bounds. 

To be a good candidate for a black hole mimicker, any such object must concentrate its mass in a
radius $R_s \gtrsim 2 G_N M$, while avoiding the formation of the event horizon and of the singularity.
Based on these and other considerations \cite{Frolov_1990, Dymnikova_1992, thooft_1998, Chapline_2003, Chapline_2005}, Mazur and Mottola proposed the model of a
``gravitational vacuum star'', also called gravastar \cite{Mazur_2023, Mottola_2023}. As a star undergoes gravitational collapse, the quantum vacuum undergoes a phase
transition at or near the location where the event horizon is expected to form, similar to the
quantum liquid-vapor critical point of an interacting Bose fluid \cite{Chapline_2003, Chapline_2005}. The interior of the
critical surface at the horizon is sustained by a fluid with negative pressure $p = - \rho$, separated from the exterior Schwarzschild metric by a shell of material with  equation of state $p = + \rho$. 

For static and spherically symmetric solutions the exterior metric of a gravastar is the same as that of a black hole down to the length scale of the shell. Therefore, it is very difficult to tell them apart experimentally. For example, since the radius of a gravastar is arbitrarily close to its Schwarzschild radius, the light it emits will be largely redshifted \cite{Abramowicz_2002}, to the point that a gravastar is essentially indistinguishable from a black hole if we look at electromagnetic radiation only.

Nonetheless, there are several proposed observational tests to differentiate between these objects \cite{Cardoso_2019}. For example, in \cite{Chirenti_2007} the stability against axial perturbations was studied and it was found that the eigenfrequencies of quasinormal modes indeed differ for the two objects. Although the literature on these observational tests is extensive and growing larger \cite{Broderick_2007, Pani_2009, cardoso1, cardoso2, Rosa_2024}, as of now the question regarding the existence of gravastars remains unanswered.

In this work, we will add to this line of research by showing that the exterior metric of a gravastar differs from that of a black hole, once the leading corrections from quantum gravity are taken into account. In particular, we will focus on the unique effective action of quantum gravity developed by Vilkovisky and Barvinsky \cite{Barvinsky_1983, Barvinsky_1985, Barvinsky_1987, Barvinsky_1990}, for which it has been pointed out that the non-locality of this action gives rise to quantum gravitational hair \cite{Calmet_2022qh, Calmet_2022, Calmet_2023}. More precisely, we will show that the exterior geometry contains information about the equation of state in the interior geometry. In contrast to earlier examples within this formalism, it turns out that this type of quantum gravitational hair appears already at order $r^{-3}$ in the asymptotic expansion. 

This feature will in principle allow to distinguish various types of dark energy stars that differ by their equation of state. Moreover, since Schwarzschild black holes do not receive quantum corrections at second order in curvature \cite{Calmet_2017qqa, Calmet_non_local_GB}, this opens the door to a set of new measurements to distinguish black holes from gravastars. As metric components are not observable by themselves, we will also translate these metric corrections into corrections to observables in the field of gravitational lensing \cite{goldstein, Evans_1996, lensing_review, photon_rings}, finding deviations from classical results.

This paper is organized as follows. In Section \ref{gravastars_section}, we introduce the gravastar and dark energy star models. In section \ref{effective_section}, we briefly review the quantum gravity effective action and the modified Einstein field equations. In Section \ref{quantum_section}, we find perturbative solutions to the equations of motion, showing the presence of quantum hair in gravastars and dark energy stars. Finally, in Section \ref{lensing_section}, we compute the quantum corrected photon sphere radius of gravastars and dark energy stars and also the angle by which light rays passing near these objects are bent. We reserve Section \ref{conclusions_section} for the conclusions.

\section{Gravastars and dark energy stars} \label{gravastars_section}
In the model proposed by Mazur and Mottola \cite{Mazur_2023, Mottola_2023}, gravastars are composed of three distinguished regions, with two infinitesimally thin layers at the junction surfaces.

The innermost region is described by the de Sitter metric:
\begin{equation}
	ds^2 = (1- H^2 r^2) dt^2 -  (1- H^2 r^2)^{-1} dr^2 - r^2 d\Omega^2, \label{deSitter}
\end{equation}
with
\begin{equation}
	H^2 =\frac{8 \pi G_N \rho}{3} = \frac{\Lambda}{3},
\end{equation}
where $\Lambda$ is the cosmological constant and the horizon is located at $r=H^{-1}$. This core is characterized by an equation of state with negative pressure $p = -\rho$, which counteracts the gravitational pull and removes the singularity at the origin, since the strong energy condition is not satisfied \cite{Hawking_1970}.

The exterior region is assumed to be vacuum, and thus, due to Birkhoff's theorem, described by the Schwarzschild metric:
\begin{equation}
	ds^2 = \left(1 - \frac{2 G_N M}{r} \right) dt^2 - \left(1 - \frac{2 G_N M}{r} \right)^{-1}dr^2 - r^2 d\Omega^2, \label{schwarzschild}
\end{equation}
where $M$ is the total ADM mass of the star.

The third region is given by a thick shell of material acting as a boundary between the interior and exterior regions. The material has equation of state $p = + \rho$ and is located at a position such that we avoid the formation of both the Schwarzschild $( R_H = 2 G_N M)$ and de Sitter $(R_H = H^{-1})$ horizons.

Several variations of gravastars have been extensively studied throughout the years \cite{Cattoen_2005, Moruno_2012, Mazur_2015}. 
In particular, a simplified model was proposed by Visser and Wiltshire \cite{Visser_2004}, where the thick shell of matter and the two junction surfaces are combined in a single infinitesimally thin shell.

In this work, we will be interested in a particular extension of the single thin-shell gravastar model which gives rise to the concept of \textit{dark energy stars} \cite{Chapline_2005, Lobo_2006}. In this model, the de Sitter interior is generalized to a region governed by an equation of state 
\begin{equation}
	p = \omega \rho, \quad \text{with} \quad \omega < -1/3.
\end{equation}
The motivation for this model comes from the observed accelerated expansion of the universe \cite{Riess_1998, Perlmutter_1999, Riess_2001, Peebles_2003, wmap, refId0}, which suggests the existence of a cosmic fluid parameterized by an equation of state with $ \omega < -1/3$. We note that current observations suggest a value of $\omega$ close to $-1$, in which case the model reduces to the gravastar model discussed above.

\begin{figure}
	\centering \hspace{1cm}
	\includegraphics[scale=0.7]{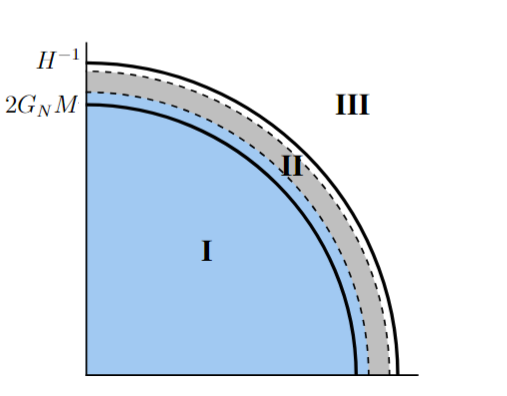}
	\caption{Representation of the three-layer dark energy star model. Region I in blue is the interior dark energy region ($p = \omega \rho, \ \omega<-1/3$), region II in gray and delimited by the two dashed lines is the thin shell ($p = + \rho$), region III in white is the exterior Schwarzschild region ($p = \rho = 0$). The thick lines correspond to the Schwarzschild and de Sitter horizons. The star radius is such that $2 G_N M \lesssim R_s \lesssim H^{-1}$.}
	\label{fig_gravastar}
\end{figure}

Summarizing, we consider a three layer star, cf. Fig. \ref{fig_gravastar}, with:
\begin{itemize}
	\item[I.] an interior dark energy region, with equation of state $p = \omega \rho$ and $\omega < -1/3$;
	\item[II.] a single thin shell $p = +\rho$, with a radius $R_s$ such that $2 G_N M \lesssim R_s \lesssim H^{-1}$, in order to avoid the formation of the event horizon;
	\item[III.] an exterior Schwarzschild region, $\rho = p = 0$.
\end{itemize}
We take the interior energy density to be homogeneous. The total ADM mass $M$ of the star is given not only by the de Sitter vacuum but also receives a contribution from the thin shell. Therefore, we parameterize the energy density as $\rho = k \rho_0$, where $\rho_0 = \frac{3 M}{4 \pi R_s^3}$ and $k \lesssim1$. The interior metric is thus \cite{Lobo_2006}
\begin{equation}
	ds^2 = \left(1 - \frac{2 G_N k M}{R_s^3} r^2 \right)^{-(1+3\omega)/2} dt^2 - \left(1 - \frac{2 G_N k M}{R_s^3} r^2  \right)^{-1} dr^2 - r^2 d\Omega^2. \label{darkstar}
\end{equation}

We can now proceed to compute the quantum corrected metric components of the dark energy star.

\section{Quantum gravity effective action} \label{effective_section}
General relativity is perturbatively non-renormalizable \cite{tHooft_1974, Goroff_1986}. However, at energy scales below the Planck mass $M_p = 2.4 \times 10^{18}$ GeV, one can use the effective field theory to make generic quantum gravity predictions. Here, we will employ the unique effective action of quantum gravity developed by Barvinsky and Vilkovisky \cite{Barvinsky_1983, Barvinsky_1985, Barvinsky_1987, Barvinsky_1990}. This effective action is obtained by integrating out the quantum fluctuations of the graviton and possibly other massless fields, and it is given by \begin{equation}
	\Gamma =\Gamma_m + \Gamma_L + \Gamma_{NL}.
\end{equation}
The first term is the usual matter action:
\begin{equation}
	\Gamma_m = \int d^4 x \sqrt{-g} \ \mathcal{L}_m.
\end{equation}
The local term, at second order in curvature, reads
\begin{multline}
	\Gamma_L = \int d^4 x \sqrt{-g} \left[ \frac{M_p^2}{2} R + c_1(\mu) R^2 + c_2(\mu) R_{\mu \nu}R^{\mu \nu} 
	+ c_3(\mu) R_{\mu \nu \alpha \beta} R^{\mu \nu \alpha \beta} + \mathcal{O}(M_p^{-2}) \right],  \label{local_action}
\end{multline}
where the prefactors $c_i$ are the Wilson coefficients and $\mu$ is the renormalization scale. The non-local part is
\begin{multline}
	\Gamma_{NL} = - \int d^4 x \sqrt{-g} \left[\alpha R \ln\left(\frac{\square}{\mu^2} \right)R + \beta R_{\mu \nu} \ln\left(\frac{\square}{\mu^2} \right) R^{\mu \nu} 
	\right. \\ \left.
	+\gamma R_{\mu \nu \alpha \beta}\ln\left(\frac{\square}{\mu^2} \right)R^{\mu \nu \alpha \beta} + \mathcal{O}(M_p^{-2}) \right],
\end{multline}
where $\square \coloneqq g_{\mu \nu} \nabla^\mu \nabla^\nu$. A discussion of the non-local operator $\ln(\square/\mu^2)$ acting on radial functions can, for example, be found in \cite{Calmet:2018rkj,Calmet_2019}.

The last term of the local action (\ref{local_action}), containing the contraction of two Riemann tensors, can be rewritten as a function of the Ricci tensor and Ricci scalar using the Gauss-Bonnet topological invariant:
\begin{equation}
	\int d^4 x \sqrt{-g} (R^2 - 4 R_{\mu \nu} R^{\mu \nu} + R_{\mu \nu \alpha \beta} R^{\mu \nu \alpha \beta}) = 32 \pi^2 \chi(\mathcal{M}), 
\end{equation}
where $\chi(\mathcal{M})$ is the Euler characteristic of the manifold. Being a topological term, it does not affect the equations of motion. 
In this way we may simplify the local action to
\begin{equation}
	\Gamma_L = \int d^4 x \sqrt{-g} \left[ \frac{R}{16 \pi G_N} + \Bar{c}_1 R^2 + \Bar{c}_2 R_{\mu \nu}R^{\mu \nu}\right],
\end{equation}
where $\Bar{c}_1=c_1-c_3$ and $\Bar{c}_2=c_2+ 4 c_3$.

The values of the Wilson coefficients $c_i$ of the local part depend on the UV completion of quantum gravity, and follow from a matching of the effective action to the UV-completion, see e.g. \cite{Calmet:2024neu}. The values of the non-local part are instead independent of the UV-completion \cite{Barvinsky_1983, Barvinsky_1985, Donoghue_2014} and can be calculated in a gauge invariant manner. The values for various types of matter are listed in Tab. \ref{wilson}. Denoting by $N_s$, $N_f$, $N_v$, $N_g$ the number of scalar, fermionic, vector and graviton fields in the theory, we have in general
\begin{equation}
	\alpha = N_s \alpha_s + N_f \alpha_f + N_v \alpha_v + N_g \alpha_g \, .
\end{equation}

\begin{table}
	\centering
	\begin{tabularx}{\textwidth}{|>{\centering\arraybackslash}X|>{\centering\arraybackslash}X|>{\centering\arraybackslash}X|>{\centering\arraybackslash}X|}
		\hline
		& $\alpha$ & $\beta$ & $\gamma$ \\
		\hline
		Scalar & $5(6\xi - 1)^2$ & $-2$ & $2$ \\
		Fermion & $-5$ & $8$ & $7$ \\
		Vector & $-50$ & $176$ & $-26$ \\
		Graviton & $250$ & $-244$ & $424$ \\
		\hline
	\end{tabularx}
	\caption{Non-local Wilson coefficients for different fields. All numbers should be divided by $11520 \pi^2$. $\xi$ is the value of the non-minimal coupling for a scalar theory.}
	\label{wilson}
\end{table}

By varying the effective action with respect to the metric, one obtains the equations of motion, see e.g. \cite{Calmet_2019},
\begin{equation}
	G_{\mu \nu}  + 16 \pi G_N (H_{\mu \nu}^L + H_{\mu \nu}^{NL}) = 8 \pi G_N T_{\mu \nu} \, ,
	\label{lin_einstein}
\end{equation}
where
\begin{equation}
	G_{\mu \nu} = R_{\mu \nu} - \frac{1}{2}R g_{\mu \nu}
\end{equation}
is the usual Einstein tensor and $T_{\mu \nu}$ the energy-momentum tensor. The local part is given by
\begin{equation}
	\begin{split}
		H_{\mu \nu}^L=&\Bar{c}_1 \left(2 R R_{\mu \nu} - \frac{1}{2} g_{\mu \nu} R^2 +2 g_{\mu \nu} \square R-2 \nabla_\mu \nabla_\nu R \right) \\ &+ \Bar{c}_2 \left(2{R^{\alpha}}_{\mu} R_{\nu \alpha} - \frac{1}{2}g_{\mu\nu}R_{\alpha\beta}R^{\alpha\beta} + \square R_{\mu\nu} + \frac{1}{2}g_{\mu\nu}\square R - \nabla_{\alpha}\nabla_{\mu}{R^{\alpha}}_{\nu}  - \nabla_{\alpha}\nabla_{\nu}{R^{\alpha}}_{\mu}
		\right), \label{local}
	\end{split}
\end{equation}
and the non-local part is
\begin{equation}
	\begin{split}
		H_{\mu \nu}^{NL}=&-2 \alpha \left(R_{\mu\nu} - \frac{1}{4}g_{\mu\nu}R + g_{\mu\nu}\square - \nabla_{\mu}\nabla_{\nu} \right) \ln \left(\frac{\square}{\mu^2} \right)R  \\ &-\beta \left(2\delta^{\alpha}_{(\mu} R_{\nu)\beta} - \frac{1}{2}g_{\mu\nu}{R^{\alpha}}_{\beta} + \delta^{\alpha}_{\mu}g_{\nu\beta}\square + g_{\mu\nu}\nabla^{\alpha}\nabla_{\beta} - \delta^{\alpha}_{\mu}\nabla_{\beta}\nabla_{\nu} - \delta^{\alpha}_{\nu}\nabla_{\beta}\nabla_{\mu}
		\right)\ln \left(\frac{\square}{\mu^2} \right) {R^\beta}_\alpha \\
		&-2\gamma \left( \delta^{\alpha}_{(\mu} R_{\nu)\sigma \tau}^{\beta} - \frac{1}{4}g^{\mu\nu}{R^{\alpha\beta}}_{\sigma\tau} + (\delta^{\alpha}_{\mu}g_{\nu \sigma} +\delta^{\alpha}_{\nu}g_{\mu \sigma} ) \nabla^{\beta}\nabla_{\tau}
		\right)\ln \left(\frac{\square}{\mu^2} \right) {R_{\alpha \beta}}^{\sigma \tau}. \label{nonlocal}
	\end{split}
\end{equation}
Note that variations of the $\ln (\square / \mu^2)$ terms yield terms of higher order in curvature which can then be neglected at second order in the curvature expansion \cite{Donoghue_2015}. 

We solve the equations of motion (\ref{lin_einstein}) perturbatively in the Planck length. That is, we consider perturbations of the above metrics of the form
\begin{equation}
	\Tilde{g}_{\mu \nu} = g_{\mu \nu} + h_{\mu \nu},
\end{equation}
where $g_{\mu \nu}$ is the classical background metric and the perturbation $h_{\mu \nu}$ is taken to be of order $\mathcal{O}(l_p^2)$. Additionally, we linearize the equation, such that the equations of motion (\ref{lin_einstein}) become
\begin{equation}
	G_{\mu \nu}^L [h] + 16 \, \pi \, l_p^2\,(H_{\mu \nu}^L[g]+H_{\mu \nu}^{NL}[g]) = 0, \label{perturb_einstein}
\end{equation}
where the linearized Einstein tensor is given by
\begin{multline}
	2\, G_{\mu \nu} ^L = \square h_{\mu \nu} - g_{\mu \nu}\square h + \nabla_\mu \nabla_\nu h - \nabla_\mu \nabla^\beta h_{\nu \beta} - \nabla_\nu \nabla^\beta h_{\mu \beta} \\ + g_{\mu \nu} \nabla^\alpha \nabla^\beta h_{\alpha \beta} + 2 {{{R^{\alpha}}_\mu}^\beta}_\nu h_{\alpha \beta}, \label{lin}
\end{multline}
and $H_{\mu \nu}^L[g]$ and $H_{\mu \nu}^{NL}[g]$ are given, respectively, by (\ref{local}) and (\ref{nonlocal}). Once we have chosen a given background we can solve for the perturbation.

\section{Quantum hair in dark energy stars} \label{quantum_section}
We impose that the perturbation is spherically symmetric and time-independent as the background metric and use the gauge freedom to set $h_{\theta \theta} = 0$. We then find  corrections $h_{\mu \nu} = \delta g_{\mu \nu}^{\text{int}}$ to the interior metric (\ref{darkstar}) given by 
\begin{align}
	\delta g_{tt}^{\text{int}} &= \left[\alpha + \beta + 3\gamma - 3 \omega (\alpha-\gamma)\right] \frac{192 \, \pi \, k \, l_p^2\, G_N M}{R_s^3} \ln \left( \frac{R^2_s}{R^2_s - r^2} \right) \nonumber \\& \quad+ \frac{C_1}{r} + C_2 + l_p^2 \, \mathcal{O}(G_N^2 M^2) + \mathcal{O}(l_p^4), \\ \delta g_{rr}^{\text{int}} &= \left[(\alpha -\gamma) - \omega (3 \alpha + \beta + \gamma)\right] \frac{384 \pi \, k \, l_p^2 \, G_N M \, r^2}{R_s^3 (R_s^2-r^2)}+ \frac{C_1}{r} + l_p^2 \, \mathcal{O}(G_N^2 M^2) + \mathcal{O}(l_p^4)\,,
\end{align}
where the integration constants $C_i$ must be set to zero if we require regularity at the origin $r = 0$. Moreover, $\mathcal{O}(l_p^4)$ terms are due to the cut-off of the effective action, whereas $l_p^2 \, \mathcal{O}(G_N^2 M^2)$ terms come from linearizing the equations of motion.

By a similar calculation we obtain corrections $h_{\mu \nu} = \delta g_{\mu \nu}^{\text{ext}}$ to the exterior metric (\ref{schwarzschild}) with
\begin{align}
	\delta g_{tt}^{\text{ext}} &= \left[\alpha + \beta + 3\gamma - 3 \omega (\alpha-\gamma)\right] \frac{192 \, \pi\, k \, l_p^2 \, G_N M}{R_s^3} 
	\left[ 2 \frac{R_s}{r} + \ln \left( \frac{r - R_s}{r + R_s} \right) \right] \nonumber \\
	&\quad + \frac{C_3}{r} + C_4 + l_p^2 \, \mathcal{O}(G_N^2 M^2) + \mathcal{O}(l_p^4) \, , \\ 
	\delta g_{rr}^{\text{ext}} &= \left[(\alpha - \gamma) - \omega (3 \alpha + \beta + \gamma)\right] \frac{384 \, \pi \, k \, l_p^2 \, G_N M}{r (r^2 - R_s^2)} 
	+ \frac{C_3}{r} + l_p^2 \, \mathcal{O}(G_N^2 M^2) + \mathcal{O}(l_p^4) \, ,
\end{align}
where the integration constants $C_i$ must be set to zero if we require asymptotic flatness, that is $\lim_{r \rightarrow \infty} \delta g_{\mu \nu}^{\text{ext}} =0$.

Far away from the star, that is for $r \gg R_s$, the exterior metric corrections reduce to
\begin{align}
	\delta g_{tt}^{\text{ext}} &= -\left[\alpha + \beta + 3\gamma - 3 \omega (\alpha - \gamma)\right] \frac{128 \,  \pi \, k \, l_p^2 \, G_N M}{r^3} + l_p^2 \, \mathcal{O}(G_N^2 M^2) + \mathcal{O}(l_p^4) \, , \label{ext_far}\\
	\delta g_{rr}^{\text{ext}} &= \left[(\alpha - \gamma) - \omega (3 \alpha + \beta + \gamma)\right] \frac{384 \, \pi\,k \, l_p^2\, G_N M}{r^3} + l_p^2 \, \mathcal{O}(G_N^2 M^2) + \mathcal{O}(l_p^4) \, ,\label{ext_far2}
\end{align}
whereas deep inside the star, that is for $r \ll R_s$, the interior corrections vanish at this order:
\begin{equation}
	\delta g_{tt}^{\text{int}} =  \delta g_{rr}^{\text{int}} = l_p^2 \, \mathcal{O}(G_N^2 M^2) + \mathcal{O}(l_p^4) \, .
\end{equation}

We note that the corrections diverge in the limit $\abs{r-R_s} \rightarrow 0^+$, which is a consequence of applying higher differential operators to a solution that is itself only once continuously differentiable. For a more detailed discussion about these divergences, we refer to \cite{Calmet_2019}, where it is shown that the above metric corrections should be considered only outside a layer of thickness $\epsilon \gtrsim l_p$ around the star surface. As a drawback, the above corrected metric cannot be used to study the stability of the model in the Israel–Lanczos–Sen junction condition formalism \cite{Israel, Lanczos, Sen}, which aims to find the equilibrium position of the freely moving transition layer at $R_s$, as is often done for gravastars. 
\par 

In this work, we focus on the region $r\gg R_s$ observed by an asymptotic observer. Taking the corrections \eqref{ext_far} and \eqref{ext_far2} as a proxy for the entire geometry, the asymptotic observer will find a shifted horizon radius, which may affect the model as discussed in the previous section. The gravitational radius $R_H$ can be found by solving the condition
\begin{equation}
	g^{rr}(R_H) = 0 \, .
\end{equation}
For our case this implies
\begin{equation}
	\begin{split}
		r - 2 \, G_N M = -\frac{384 \, \pi \, k \, l_p^2\, G_N M [(\alpha -\gamma) - \omega (3 \alpha + \beta + \gamma)]}{r^2} \label{schwarz_horizon_eq} \, .
	\end{split}
\end{equation}
We solve this equation perturbatively, which yields the shifted radius
\begin{equation}
	R_H = 2 \, G_N M - \frac{96 \, \pi \, k\, l_p^2}{G_N M} \, \Big[(\alpha -\gamma) - \omega \, (3 \alpha + \beta + \gamma)\Big] \, .
\end{equation}   
The extra terms are subleading with respect to the classical result and will not affect the basic features of the model described in figure \ref{fig_gravastar}. We emphasize, however, that this analysis is performed using the asymptotic expansion for which the linearized Einstein equations provide a good approximation. It is thus difficult to make definite statement about the stability of the model within our approximation.
\par 
We highlight the important result that all of the corrections depend on the parameter $\omega$ of the equation of state. This is a clear example of a quantum gravitational hair: an outside observer can recover information about the equation of state of the interior fluid by probing metric corrections in the weak field region. This confirms that quantum hair is a generic feature of the effective action of quantum gravity \cite{Calmet_2022qh, Calmet_2022, Calmet_2023}. Remarkably, this type of hair, that is the dependence on the equation of state of the star, appears already at order $(l_p^2 G_N M )/ r^3$. In previous works, on the other hand, the quantum corrections appeared at the same order $(l_p^2 G_N M )/ r^3$, but the hair, that is the dependence on the density of the star, was only apparent at order $( l_p^2 G_N M R_s^2)/ r^5$. Note that the latter type of hair can also be found for the dark energy stars discussed in this paper, if different density profiles are studied. However, it will be subleading compared to the former type of hair, i.e. the dependence on the equation of state.
\par 

From an observational perspective this feature is potentially interesting, as it allows to distinguish dark energy stars with a different equation by evaluating the quantum corrections to the gravitational potential at infinity. We note that for $\omega = 0$ we recover the corrections to the ball of dust studied in \cite{Calmet_2019}, as expected. Similarly, for $\omega=-1$, one obtains the gravastar model. Since there are no quantum corrections to the potential for a Schwarzschild black hole at this order in the expansion \cite{Calmet_2017qqa, Calmet_non_local_GB}, this would also allow to distinguish gravastars from black holes by measuring the quantum corrections to the weak field metric.

\section{Observables in gravitational lensing} \label{lensing_section}
Metric components are not measurable by themselves, therefore we will now calculate the induced corrections to observable quantities. Here, we will be particularly interested in the quantum corrections to gravitational lensing, that is the collection of all the effects caused by a gravitational field on the propagation of electromagnetic radiation.

\subsection{Photon sphere}
Motivated by the images of the black holes Sgr A* at the center of the Milky Way \cite{sgr1,sgr2,sgr3,sgr4,sgr5,sgr6} and M87* at the center of the galaxy Messier 87 \cite{M87_1, M87_2, M87_3, M87_4}, we start by analyzing the quantum correction to the photon spheres \cite{photon_rings}, that is photons moving along the unstable circular orbit around the source. 

Let us write the generic line element for a static and spherically symmetric metric as
\begin{equation}
	ds^2 = f(r) dt^2 - g(r)^{-1} dr^2 - r^2 (d \theta^2 + \sin^2 \theta d\phi^2). \label{general_metric}
\end{equation}
The two Killing vectors $\Vec{k} = \Vec{\partial_t}$ (associated to time translation invariance) and $\Vec{n}=\Vec{\partial_\phi}$ (associated to rotations around the $z$-axis)
imply the existence of the two integrals of motion
\begin{align}
	E &= - k_\mu u^\mu = f(r) \frac{dt}{d\lambda} \, , \\
	L &= n_\mu u^\mu = r^2 \sin^2 (\theta) \frac{d \phi}{d \lambda} \, , \label{L_integral_of_motion}
\end{align}
where $u^\mu \equiv dx^\mu/d\lambda$ is the photon four-momentum.
Without loss of generality, we can restrict the motion to be on the equatorial plane $\theta= \pi/2$. Using then the condition $g_{\mu \nu} u^\mu u^\nu = 0$, we find
\begin{equation}
	\frac{f(r)}{g(r)} \left(\frac{dr}{d\lambda}\right)^2 + V(r, E, L) = 0 \, , \label{photon_sphere_equation}
\end{equation}
where the effective potential is given by
\begin{equation}
	V(r, E, L)=f(r) \frac{L^2}{r^2}-E^2 \, .
\end{equation}
Circular orbits are obtained from
\begin{equation}
	\frac{dr}{d\lambda}=\frac{d^2r}{d\lambda^2}=0 \, ,
\end{equation}
which translates into the conditions for the potential
\begin{equation}
	V(r_p)=V'(r_p)=0 \, . \label{photon_sphere-conditions}
\end{equation}
This defines the so called \textit{photon sphere} at $r_p$ that gives raise to a gravitational lensing generating infinitely-many images. Solving the first equation (\ref{photon_sphere-conditions}) for the impact parameter $b \equiv L/E$ and then plugging it into the second equation we find that the latter is satisfied when
\begin{equation}
	f'(r_p) r_p - 2 f(r_p)=0 \, .
\end{equation}
It can be shown that any spherically symmetric and static spacetime with an horizon at $r=R_H$ and which is asymptotically flat must have a light sphere at a radius between the horizon radius and infinity \cite{photon_sphere}. For the classical Schwarzschild metric (\ref{schwarzschild}) this is located at
\begin{equation}
	r_p = 3 G_N M \, .
\end{equation}

For the quantum corrected dark energy star we have
\begin{equation}
	f(r) = 1 - \frac{2 G_N M}{r} -\frac{128 \, \pi \, k \, l_p^2\, G_N M}{r^3} \left[\alpha + \beta + 3\gamma - 3 \omega (\alpha - \gamma)\right],  
\end{equation} 
leading to
\begin{equation}
	r - 3 \, G_N M = \left[\alpha + \beta + 3\gamma - 3 \,\omega \, (\alpha - \gamma)\right] \frac{ 320 \, \pi\, k\, l_p^2 \, G_N M}{r^2} \, .
\end{equation}
We can solve this equation perturbatively around the classical result $r_p = 3 G_N M$ and obtain the modified photon sphere radius
\begin{equation}
	r_p = 3 G_N M + \Big[\alpha + \beta + 3 \gamma -3\omega(\alpha - \gamma) \Big]  \, \frac{ 320 \, \pi\,  k\, l_p^2}{9\, G_N M} \, .
\end{equation}
We emphasize that the above shift depends on the parameter $\omega$, allowing, in principle, to observationally distinguish dark energy stars with different equations of state. Moreover, as black holes do not receive corrections at second order in $G_N$, the location of the photon sphere also allows to distinguish gravastars from black holes.

\subsection{Bending of light rays}
Light rays passing near a massive object will be bent by an angle $\phi$ with respect to their original trajectory. Using the metric (\ref{general_metric}) and working on the equatorial plane, we may rewrite the condition that light rays move along null geodesics, i.e. $g_{\mu \nu} u^\mu u^\nu = 0$, as \cite{Evans_1996}
\begin{equation}
	\frac{d^2 u}{d \phi^2} + u f(u) + \frac{u^2}{2}  \frac{df}{du} = 0,
\end{equation}
where $ u = r^{-1}$. For the classical Schwarzschild metric we find
\begin{equation}
	\frac{d^2 u}{d \phi^2} + u = 3 G_N M u^2. \label{diff_lens}
\end{equation}
We solve this differential equation perturbatively, i.e. we first set the right-hand side to zero and obtain the zeroth-order solution
\begin{equation}
	u = \frac{\sin \phi}{R}, \label{zeroth_order_solution}
\end{equation}
where $R$ is the distance of closest approach to the origin. We can then plug this result in the right-hand side of (\ref{diff_lens}) and solve to obtain the first-order solution
\begin{equation}
	u = \frac{\sin \phi}{R} + \frac{3 G_N M}{2 R^2} \left(1+\frac{1}{3}\cos 2\phi \right) . \label{u_classic}
\end{equation}
At large distances $r \rightarrow \infty$, $u \rightarrow 0$ and the deflection angle becomes small $\phi \rightarrow \phi_\infty$. We can thus expand (\ref{u_classic}) as
\begin{equation}
	0 = \frac{\phi_\infty}{R} + \frac{2 G_N M}{R^2}, \label{deflection_angle}
\end{equation}
and the total deflection $\Delta \phi_\infty = 2 \abs{\phi_\infty}$ is
\begin{equation}
	\Delta \phi_\infty = \frac{4 G_N M}{R} ,
\end{equation}
which is the very well known result of general relativity.

Repeating the calculation for the quantum corrected dark energy star, in the limit $r \gg R_s$, we find the total deflection angle
\begin{equation}
	\Delta \phi_\infty = \frac{4 G_N M}{R} + \frac{1024 \, \pi\, k \, l_p^2 \, G_N M[\alpha + \beta + 3\gamma - 3\omega (\alpha-\gamma)]} {3 R^3}.
\end{equation}
As the result depends on the parameter $\omega$, this deviation can also be used to observationally distinguish dark energy stars with different equations of state, and to distinguish gravastars from Schwarzschild black holes.

\section{Conclusions} \label{conclusions_section}
In this letter, we used the effective action of quantum gravity at second order in curvature to compute corrections to the metric of gravastars and dark energy stars. This gives rise to the presence of hair in the leading quantum gravitational corrections to these models, showing explicitly how the corrections to the exterior metric depend on the equation of state of the interior fluid. 
\par 

We then proceeded to study implications of the metric corrections for the gravitational lensing, analysing how the modified metrics affect the radius of the photon sphere and the deflection angle of light rays passing near a gravitational lens. We found that these quantities deviate from the classical results by terms of order $\mathcal{O}(l_p^2 G_N M)$. The fact that these deviations depend on the equation of state parameter $\omega$ allows to observationally distinguish dark energy stars with different equations of state from classical black holes.
\par 

It must be noted, however, that these effects are Planck scale suppressed with respect to the classical results and therefore undetectable with current technology. It would thus be interesting to investigate the possibility of the accumulation of such corrections, for example over cosmological time scales. The study of such an accumulation will likely require to go beyond the linear approximation employed in this work.
\par 

Moreover, we note that the results apply only to the case of non-rotating objects, whereas the exterior of astrophysical black holes is better described by the Kerr metric. Extending our results to this more general case is a non-trivial exercise, which requires a better understanding of the $\ln (\square / \mu^2)$ operator on Kerr spacetimes. Furthermore, the equations that need to be solved for finding the components of the perturbation $h_{\mu \nu}$ in the rotating case will form a system of coupled differential equations both in the radial and angular coordinate. In future work, it would be interesting to see if the quantum corrections for rotating objects can become comparable to the classical results, in the case of very large angular momentum. Nonetheless, the results for the non-rotating objects studied in this paper will still apply as a leading order (in angular momentum) approximation to the case of a slowly rotating black hole. 
\par 

Despite these two limitations, we conclude that the results presented in this work provide a very clear example of quantum gravitational hair that is independent of the UV-completion of quantum gravity. Moreover, the presence of this hair could not only help in tackling theoretical questions \cite{Calmet:2021cip,Calmet:2024tgm}, but also opens up a new avenue towards distinguishing gravastars from black holes. This could thus shed a new light on the viability of gravastar models as alternative to black holes.

\section*{Acknowledgements}

We are grateful to Xavier Calmet for various useful discussions. The work of FK is supported by a postdoctoral fellowship of the Alexander von Humboldt foundation and a fellowship supplement of the Carl Friedrich von Siemens foundation. IP would like to thank the Arnold Sommerfeld Center for Theoretical Physics of the Ludwig-Maximilians-Universit{\"a}t of Munich for hosting him during the completion of this research.
RC is partially supported by the INFN grant FLAG and his work has also been carried out in the framework of activities of the National Group of Mathematical Physics (GNFM, INdAM). RC acknowledges the support of PNRR MUR project PE0000023-NQSTI (Italy) financed by the European Union – Next Generation EU.

\printbibliography[heading=bibintoc, title={Bibliography}]

\end{document}